\documentclass[pdflatex,sn-basic]{sn-jnl}% Springer basic style, author-year citations

\usepackage{graphicx}%
\usepackage{multirow}%
\usepackage{amsmath,amssymb,amsfonts}%
\usepackage{amsthm}%
\usepackage{xcolor}%
\usepackage{textcomp}%
\usepackage{booktabs}%
\usepackage{listings}%
\usepackage{array}%
\usepackage{tabularx}%
\usepackage[UKenglish]{babel}%

\usepackage{xspace}
\xspaceaddexceptions{\,}

\newcommand\eg{e.g.\@\xspace}

\newcommand{\uwcet}[1]{\textbf{#1}}
\newcommand\llta{\textsc{llta}\xspace}
\newcommand\llvmta{\textsc{llvmta}\xspace}
\newcommand\llvm{\textsc{llvm}\xspace}
\newcommand\otawa{\textsc{otawa}\xspace}
\newcommand\aiT{aiT\xspace}
\newcommand\heptane{Heptane\xspace}
\newcommand\prepphase{Preparation Phase\xspace}
\newcommand\anaphase{Analysis Phase\xspace}

\usepackage{tikz}
\usetikzlibrary{arrows,
	arrows.meta,
	backgrounds,
	calc,
	chains,
	decorations,
	decorations.pathreplacing,
	decorations.pathmorphing,
	decorations.markings,
	fit,
	matrix,
	patterns,
	positioning,
	shadows,
	shapes}
\usepackage{pgfplots}
\usepackage{pgfplotstable}
\usepackage[normalem]{ulem}

\definecolor{uds-mcyan}{RGB}{068, 119, 170}
\definecolor{uds-lcyan}{RGB}{119, 170, 221}

\newenvironment{romanenumerate}%
  {\renewcommand\theenumi{\roman{enumi}}\begin{enumerate}[(iii)]}%
  {\end{enumerate}}

\usepackage[capitalise,noabbrev]{cleveref}

\begin{document}

\title[]{LLTA: A Simplicity-Oriented Open-Source WCET Analyser}

\author*[1,2]{\fnm{Nils} \sur{Hölscher}}\email{nils.hoelscher@tu-dortmund.de}% ORCID 0000-0002-6397-5134
\author[2]{\fnm{Kay} \sur{Heider}}\email{kay.heider@tu-dortmund.de}% ORCID 0009-0005-0551-6596
\author[1,2]{\fnm{Jian-Jia} \sur{Chen}}\email{jian-jia.chen@tu-dortmund.de}% ORCID 0000-0001-8114-9760

\affil[1]{\orgname{RWTH Aachen University}, \orgaddress{\city{Aachen}, \country{Germany}}}
\affil[2]{\orgname{TU Dortmund University}, \orgaddress{\city{Dortmund}, \country{Germany}}}

\abstract{%
Deriving a safe upper bound on the worst-case execution time (WCET)
  of a real-time task is essential for hard real-time systems.  Many
  WCET analysers exist, but they are either (i) closed source or
  (ii) do not provide a WCET for an existing microcontroller.  Hence,
  it is difficult to obtain a trustworthy WCET bound for the binary
  that is flashed to an available embedded hardware platform and to
  observe whether the bound is safe.

  To solve this, we present \llta, an open-source WCET analyser for
  Commercial-Off-The-Shelf (COTS) microcontrollers. \llta is
  designed based on the following goals: it is simple (G1), runs on
  real COTS hardware (G2), reports results
  that are unsound due to programming or device constraints (G3), and
  is available open-source without licensing complications (G4).
  \llta generates WCETs for the ESP32-C6 and the MSP430(FR5994), which
  are broadly available.  This paper systematically lays out our
  design decisions for \llta with those goals in mind.  To demonstrate
  its ease of use, we propose scenarios where \llta can be used for
  lab exercises of real-time systems.
}


\maketitle

\section{Introduction}
\label{sec:introduction}
Deriving the worst-case execution time (WCET) is necessary to ensure
the timeliness of hard real-time tasks in cyber-physical systems. From
the analytical perspective, the WCET $C_i$ of a task is merely a
specified parameter. However, for hands-on engineering of real-time
systems, this is a huge hurdle and usually the WCET is just measured.
Designers, engineers, and researchers who are interested in building real-time systems should easily be able to compile their
tasks and obtain a trustworthy WCET bound for the binary they are about to flash to an embedded hardware platform.
To achieve such objectives,  Commercial-Off-The-Shelf (COTS) platforms have to be targeted, rendering the work of measuring unnecessary, which takes up more time and might not even provide safe bounds.
In the current hardware and
software ecosystem, however, it is hard to find COTS platforms for which
this loop can be closed: if a real-time tool-chain exists for a device
at all, it is usually at least partially proprietary, and devices
supported by an open-source WCET analyzer are scarce.

This scarcity is not due to the lack of static WCET research. Static
WCET analysis has been developed since decades, but none of the tools discussed here
 fits the demand of the engineering loop sketched above. The industrial analyzer
\aiT~\citep{Ferdinand2001} is sound and supports real processors, but
is closed source and priced for the certification market, which is a significant hurdle to get access to. Open research tools such as \llvmta~\citep{Hahn2022},
\heptane~\citep{Hardy2017} and \otawa~\citep{Ballabriga2010} are built
to prototype and evaluate novel WCET analysis methodologies. They target
generic or user-described processor models rather than a particular
chip one can buy, and \llvmta even declares the modelling of real-world
hardware an explicit non-goal. We discuss this landscape and how our
goals differ in \Cref{sec:goals}.

In this article we present \llta, a WCET
analyzer built exactly for simplicity and ease of use and the
centrepiece of the effort of our group towards a complete, simple and open
tool-chain for practical engineering of real-time systems. \llta is inspired by
\llvmta~\citep{Hahn2022} and, like it, implements a \emph{top-down}
approach first, where the analysis
operates as a white box inside the compiler. Concretely, \llta is a
 modification of the \llvm~\citep{Lattner2004} code generator
\textsc{llc} that injects timing-analysis passes into the ordinary
\textsc{clang}/\textsc{opt}/\textsc{llc} flow. Since a \emph{top-down} approach alone is not sufficient \llta
additionally reads the linked binary, in Executable and Linkable Format
(ELF), \emph{bottom-up} to recover the real
instruction and library-call addresses, and it decodes binary code
where a library routine is present only in the linked binary. 
Thus, \llta is less agnostic about the \emph{top-down} approach, and uses 
\emph{bottom-up} when necessary to achieve its goals and can be seen as a 
hybrid analyzer.

\llta targets two broadly
available and affordable COTS micro-controller families, the Texas
Instruments
MSP430 (MSP430FR5994) and the Espressif ESP32-C6. Both target models
are discussed in \Cref{sec:targets}: the timing
behaviour of the two devices was validated by measurements,
and their undocumented instruction caches were reverse
engineered. Restricting the targets to
simple micro-controllers whose timing needs no pipeline model keeps
both the analysis
and the tool itself small. 
Crucially for fast prototyping, \llta is honest
about its own limits: every reported bound is either sound with
respect to the measured timing models of \Cref{sec:targets} or
explicitly flagged as \emph{unsound}, \eg when the program reaches a
runtime routine whose cost \llta cannot determine.

\llta is an open-source tool for research and education and provides no guarantees for the bounds it provides. In summary, this article makes the following contributions:
\begin{itemize}
\item \llta, an open-source \llvm-based WCET analyzer designed for simplicity,
\begin{center}
  \large\url{https://github.com/tu-dortmund-ls12-rt/LLTA},
\end{center}
  which verifies its analysis against a linked binary program
  that is flashed and reports every bound as sound or explicitly
  unsound (\Cref{sec:goals,sec:llta-design});
\item measurement-validated timing models for two COTS devices, the
  MSP430FR5994 and the ESP32-C6, backed by a measurement and
  reverse-engineering effort (\Cref{sec:targets});
\item a case study on the M\"alardalen benchmark
  suite~\citep{Gustafsson2010} comparing
  \llta's bounds against cycle counts measured on both devices
  (\Cref{sec:llta-eval}).
\item a proposal to utilize \llta{} in a real-time operating systems
  (RTOS) or general real-time systems course, for students to practice
  how to derive and optimize trustworthy WCET bounds from source
  codes, in~\Cref{sec:exercises}.
\end{itemize}
The tool-chain effort \llta belongs to further
building blocks: TEACH~\citep{Teach}, an accessible case-study platform on COTS hardware
based on FreeRTOS~\citep{FreeRTOS}, a
machine-verified earliest-deadline-first (EDF) scheduler
implementation for
FreeRTOS~\citep{kuhse2026deductiveverificationearliestdeadline}, and
SLIME~\citep{hakert2026slime}, a framework that extracts task-set
models from compiled real-time operating system (RTOS) binaries.

The remainder of this article is organised as follows.
\Cref{sec:goals} states \llta's goals and contrasts its design
philosophy with existing WCET tools. \Cref{sec:background} recalls the
WCET analysis methods \llta builds on. \Cref{sec:llta-design} describes
how \llta builds, enriches and analyses its program graph.
\Cref{sec:targets} introduces the two target devices and the
measurement campaigns behind their timing models,
\Cref{sec:llta-eval} shows a case study of \llta on the M\"alardalen suite,
\Cref{sec:exercises} sketches how \llta can be used in teaching, and
\Cref{sec:conclusion} concludes the article and gives an outlook.

\section{Goals and Design Philosophy}
\label{sec:goals}

We state the goals of \llta and compare them with the goals and design
philosophies of existing WCET analyzers.

\subsection{Goals}
\label{sec:goals-list}

We target the following goals:

\textbf{Simplicity (G1):}
\llta should be small enough to be read, understood and modified
with a minimal effort. A newcomer
should not have to work through a large body of documentation before
getting to the problem at hand. This goal also dictates what \llta
does \emph{not} attempt: it only targets micro-controllers with
simple or no pipelining, because supporting complex out-of-order
micro-architectures would also require complexer WCET analysis.

\textbf{Real COTS hardware (G2):} The derived WCET bounds must hold
for the binary that is flashed onto an available and
affordable device, so that users can check them by measurement on
their own desk. The timing models behind such bounds must be
trustworthy for the real silicon rather than transcribed from
datasheets: hand-written processor models can be notoriously
error-prone~\citep{Abel2019}.

\textbf{Soundness, or an explicit warning (G3):}
In case the analysis cannot guarantee to derive a safe WCET bound, it must not silently
return a best-effort number but flag the resulting bound as
\emph{unsound}, with a warning message whenever possible. This is especially useful for the users to
tell \emph{why} a bound can fail, instead of trusting a number that happens to be wrong.

\textbf{Openness without licence restrictions (G4):}
Every component must be open source and installable: the analyzer itself, the compiler, the
cross-compilation tool chains and, deliberately, the
integer-linear-programming (ILP) solver.

\subsection{Comparison with Existing WCET Tools}
\label{sec:tool-comparison}
\llvmta~\citep{Hahn2022} is closest to \llta and its direct
inspiration. Yet, its goals are almost complementary to ours. \llvmta
aims to enable ``\emph{the evaluation of novel WCET analysis approaches in
a state-of-the-art analysis framework without dealing with the
complexity of modeling real-world hardware
architectures}''~\citep{Hahn2022}. Consequently, it models generic
textbook pipelines, from in-order to out-of-order, for the ARM
and RISC-V instruction sets, it analyses the machine-level representation under an assumed
address mapping instead of verifying it against the linked binary,
and it uses commercial solvers like
CPLEX and Gurobi. Each choice follows from that purpose, judging
whether a novel analysis idea pays off in principle, and each
conflicts with one of our goals G1--G4. This is also why we built \llta as a
new, small tool that shares only \llvmta's compiler approach with \llvm,
rather than extending \llvmta itself.

One consequence of using \llvm is worth naming, because it decides
what has to be installed. \llvmta is distributed together with a
patched version of \llvm~\citep{Hahn2022}, whereas \llta builds against an
unmodified \llvm release. A timing analysis over a whole-program graph
needs every machine function to stay alive after code generation, while
\llvm's own emit pipeline frees each one as soon as it has been
emitted. \llta meets that requirement inside its own code base, which
rebuilds the emit pipeline from public \llvm interfaces and leaves out
the pass that frees the machine functions. Hence, there is no need to modify any \llvm source file.
This is the difference between installing
a released compiler (for \llta) and a patched one (for \llvmta).

\aiT~\citep{Ferdinand2001} sits at the opposite end of the spectrum:
it is the industrial reference for sound WCET analysis of real,
complex processors. However, it is closed source and priced for the
certification market. For ease of use, where the point is precisely
to lower the hurdle of entry, a closed tool
is not an option.

\heptane~\citep{Hardy2017} comes closest to \llta in philosophy: it is
open source, usable with a free ILP solver, and deliberately
minimalistic, as its authors keep ``\emph{only a minimum number of robust
analyses}''~\citep{Hardy2017} in the main branch. The tools differ in
what their models mean. \heptane analyses MIPS and ARMv7 binaries
against a user-parameterised, idealised processor description, a
simple timing-anomaly-free pipeline with configurable caches, not
against a validated model of a specific deployable board. \llta
instead binds itself to two concrete device families.

\otawa~\citep{Ballabriga2010} is an open framework for \emph{building}
WCET analyzers, retargetable through hand-written processor
descriptions in the Sim-NML language. Among the remaining
open tools, Chronos~\citep{Li2007} implements low-level timing
analyses but only for the SimpleScalar simulation architecture, and
\textsc{sweet}~\citep{Lisper2004} focuses on flow analysis without a
hardware-level timing model.

In short, the existing open tools optimise for methodological
generality, and the industrial tool for certification-grade coverage
of complex processors. \llta focuses on determining a bound on real, affordable hardware that is either
sound or explicitly flagged as unsound, produced by a tool that is
small enough and openly available.

\section{Background}
\label{sec:background}

This section introduces the standard methods of static WCET analysis
that \llta builds upon. More detailed information can be found in the survey by
\citet{Wilhelm2008}.

A WCET bound is an upper bound on the execution time
of a program. Hence, the analysis needs one finite object
that covers all execution scenarios of a program.  This is done by
exploring the control-flow graph (CFG) of the program.  Each of its
nodes is a basic block (BB), a run of instructions that can be entered only
at its first instruction and left only at its last. Each of the CFG's
edges is a branch, call or return that can lead from one block to
another. A binary program does not carry its own CFG. A bottom-up
analyzer recovers the graph by decoding the
executable~\citep{Theiling2000}, whereas an analysis running inside the
compiler is handed the graph the compiler already
holds~\citep{Hahn2022}.

Based on the CFG, a static analyzer proceeds in three
steps~\citep{Wilhelm2008}. It first bounds how often each part of the
graph can execute, which requires an iteration bound for every loop,
since one unbounded loop results in an unbounded WCET. It
then determines the WCET of each BB on
the target hardware. It finally searches for the WCET of an execution path, that complies with the CFG and the timing
information of the BB's.

The WCET of a BB requires to analyse the
interplay of the instructions of the BB and the
hardware. Specifically, for cache, the analysis requires to analyse
which lines the cache holds when the block is entered. Cache analysis
by abstract interpretation~\citep{Alt1996} over-approximates that
state: it computes for every program point an abstract cache state
over the cache's geometry (sets, ways, line size) and replacement
policy. Its \emph{must-analysis} determines the lines that are
resident on every execution reaching the point, such that a fetch of
such a line is guaranteed to be a hit. The complementary
\emph{may-analysis} determines the lines that can be resident on some
execution, such that a fetch outside that set is a guaranteed miss.

The precision of such an analysis rises with context sensitivity. A
block's cache behaviour differs between the first iteration of its
loop and the remaining ones, and between two call sites of the same
function. Therefore, one analysis state per node in the CFG would be too coarse.
Virtual inlining and virtual unrolling (VIVU)~\citep{Martin1998} refines
the analysis by keeping one analysis state per pair of a node and
an analysis context. A context records two pieces of information.
First, it distinguishes the iterations of every enclosing bounded loop.
In principle every iteration could receive its own context, but the
analysis cost grows with the number of contexts. Usually two contexts per loop are considered: one for the first iteration,
in which the accessed lines are typically not yet cached, and one for
all remaining iterations, in which lines loaded earlier may still be
resident. Second, the context records the chain of call sites through
which the enclosing function was reached, truncated to a bounded depth.
Two calls to the same function from different places hence keep
separate analysis states. Contexts exist only inside the analysis: they
neither duplicate nodes nor change the loop bounds.

The search for the worst-case execution is carried out without
enumerating executions. The implicit path enumeration technique
(IPET)~\citep{Li1995} states it as a single ILP. It defines one variable per graph
edge that counts how often that edge is taken, flow-conservation
constraints at every node relate those counts to each other, and the loop
bounds cap the counts on the back edges. The objective function sums the block
costs weighted by their counts. The maximum over that constraint
system is the WCET bound. Pairing an abstract-interpretation phase for
the microarchitecture with an ILP for the path search is the
architecture proposed by \citet{Theiling1998}.

\section{LLTA Design}
\label{sec:llta-design}
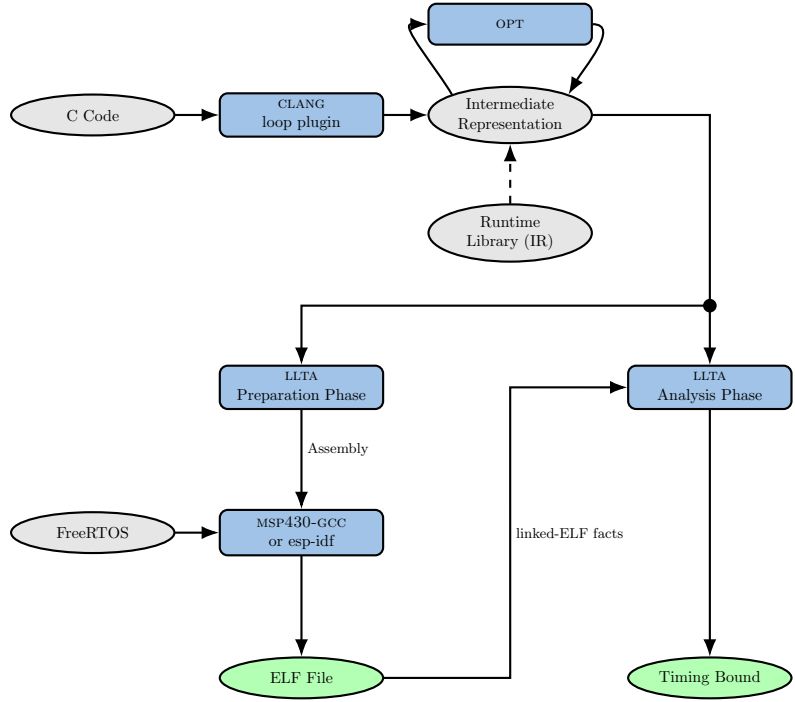
\begin{figure}[t]
	\centering
	\begin{tikzpicture}
		\begin{scope}[node distance=6mm and 9mm, thick, transform shape, scale=0.6]
			\tikzstyle{d1}=[fill=black!10]
			\tikzstyle{p1}=[fill=uds-lcyan!70]
			% every node is drawn at the same outer width: the text widths below
			% differ only to offset the ellipse's larger circumscription, and
			% minimum width governs both shapes.
			\tikzstyle{WCETall}  =[draw, text centered, rounded corners=3pt, minimum height=9mm, minimum width=36mm]
			\tikzstyle{WCETdata} =[WCETall, d1, ellipse, inner sep=2pt, text width=24mm]
			\tikzstyle{WCETphase}=[WCETall, p1, rectangle, inner sep=4pt, text width=30mm]
			% final products of the tool-chain (not intermediates)
			\tikzstyle{WCETout}  =[WCETdata, fill=green!30]

			% level 1: LLVM front-end. C Code shares the FreeRTOS column and
			% \textsc{clang} the \prepphase column.
			\node[WCETdata] (cc) at (-22mm,0) {C Code};
			\node[WCETphase] (clang) at (24mm,0) {\textsc{clang}\\ loop plugin};
			\node[WCETdata] (ir) at (70mm,0) {Intermediate Representation};
			\node[WCETphase] (opt) at (70mm,20mm) {\textsc{opt}};

			% level 2: runtime library as bitcode, ESP32-C6 only
			\node[WCETdata] (rtbc) at (70mm,-26mm) {Runtime Library (IR)};

			% level 3: the two runs of one and the same LLTA binary
			\node[WCETphase] (prep) at (24mm,-60mm) {\llta\\\prepphase};
			\node[WCETphase] (ana) at (114mm,-60mm) {\llta\\\anaphase};

			% level 4: the toolchain that links the binary
			\node[WCETphase] (linker) at (24mm,-92mm) {\textsc{msp430-gcc}\\ or esp-idf};
			\node[WCETdata] (freertos) at (-22mm,-92mm) {FreeRTOS};

			% level 5: the two products of the flow
			\node[WCETout] (bin) at (24mm,-124mm) {ELF File};
			\node[WCETout] (tb1) at (114mm,-124mm) {Timing Bound};

			% the junction where the one IR fans out to both phases, and the
			% channel that keeps the ELF edge clear of the linker column
			\coordinate (fan) at (114mm,-42mm);
			\coordinate (elfturn) at (70mm,-124mm);

			\draw[-Latex] (cc) -- (clang);
			\draw[-Latex] (clang) -- (ir);
			\draw (ir.north west) edge[-Latex,in=180,out=120] (opt.west);
			\draw (opt.east) edge[-Latex,out=0,in=60] (ir.north east);
			\draw[-Latex,dashed] (rtbc) -- (ir);

			% one intermediate representation, consumed by both phases
			\draw (ir.east) -| (fan);
			\fill (fan) circle (1.5mm);
			\draw[-Latex] (fan) -- (ana.north);
			\draw[-Latex] (fan) -- (fan -| prep.north) -- (prep.north);

			\draw[-Latex] (prep) -- node[right, pos=0.4]{\small Assembly} (linker);
			\draw[-Latex] (freertos) -- (linker);
			\draw[-Latex] (linker) -- (bin);
			\draw[-Latex] (ana) -- (tb1);
			\draw[-Latex] (bin.east) -- (elfturn) -- node[right, pos=0.5]{\small linked-ELF facts} (70mm,-60mm) -- (ana.west);
		\end{scope}
	\end{tikzpicture}
	\caption{The \llta tool flow. The ordinary \textsc{clang} and
          \textsc{opt} pipeline produces the enriched IR. \llta then runs twice
          over that same IR. The \prepphase
          emits the assembly that the target tool-chain links with FreeRTOS and device libraries, which
          makes the ELF file fully functional. The
          \anaphase receives the intermediate representation and the
          linked ELF file and computes the timing bound.
        }
	\label{fig:llta-flow}
\end{figure}

This section describes how \llta computes a WCET bound.
\Cref{fig:llta-flow} shows the flow, sketched as follows:
\begin{itemize}
\item The analysis takes the source C program as its input.
  The C program is compiled
  into \llvm intermediate representation (IR) and further utilizes the \llvm
  optimizer (\textsc{opt}), the \llvm IR linker (\textsc{llvm-link}) and our loop plugin.
  Both targets can be analysed and deployed bare metal or with FreeRTOS.
\item \llta runs twice over the \llvm IR 
  generated above:
  \begin{itemize}
  \item The \prepphase of \llta shapes the code and emits the assembly that the
    target tool chain links, detailed in
    Section~\ref{sec:llta-prep}. The assembly is further linked with FreeRTOS and device specific libraries.
    The result is a fully functional ELF binary that can be flashed onto the device.
  \item The \anaphase of \llta takes the ELF binary and the same \llvm IR to
    build an analysis graph, detailed in
    Section~\ref{sec:llta-elf}. The WCET bound is derived by further
    incorporating the corresponding costs and contexts in
    Section~\ref{sec:llta-analyses} and the Application Binary
    Interface (ABI) calls in Section~\ref{sec:abi-calls}, and the path
    analysis in Section~\ref{sec:path-analysis}.
  \end{itemize}
\end{itemize}

Whenever \llta
cannot resolve a programs WCET, it reports an explicit warning (G3).
If \llta can over-approximate but not prove the result, it emits a bound and
marks it unsound. Where the flow cannot be bounded at all, it
reports that it computed no bound. Every such refusal happens in the
\anaphase, after the \prepphase has emitted the assembly for the binary.

\subsection{Assumptions of LLTA}
\label{sec:llta-assumptions}

We assume that the WCET analysis starts from an annotated C program. \llta does not implement
its own loop bound analysis and relies on the loop bound either specified in the
C program or derived from static trip count analysis in \llvm, for example, Scalar Evolution (SCEV). Two options are
supported in \llta:
\begin{itemize}
\item Simple counted loops: \llvm's scalar-evolution analysis derives the trip
  count automatically, and \llta uses it where it succeeds.
\item For every other loop, the loop bound must be specified as part of the
  program: a \texttt{\#pragma loop\_bound(lo, hi)} annotation, written directly
  above the loop, states its iteration range, see \Cref{fig:cfg-compare} (a).
\end{itemize}
Similarly, \llta uses \texttt{\#pragma recursion\_bound} for specifying the
maximum depth of a recursion function.
\llta gives a warning about any loop that ends up without a bound, in line with goal (G3).

We further assume a timing model of the target devices consisting of a table of
per-instruction costs and a model of the instruction cache, both
measured on the real devices (\Cref{sec:targets}). Since the models are obtained by measurements, we make no claim about the
correctness of either model. An error in a model is invisible to \llta
and propagates silently into every bound computed with it.

Finally, \llta models no pipeline. Each instruction is charged its
cost from the table, and the instruction cache is the only
microarchitectural state the analysis tracks. This is what restricts
the targets to devices whose timing needs no pipeline model (G1).

\subsection{The \prepphase}
\label{sec:llta-prep}

The \prepphase{} is the first run of \llta to prepare all steps
applied to the source code for generating the corresponding ELF binary
file of the program. The goal of the \prepphase{} is to optimize the
IR code such that
\begin{romanenumerate}
\item the loop bounds can be found correctly by \llvm's analyses, making
the analyses less dependent on annotated, or wrongly annotated, loop
bounds, and
\item ABI and Library calls are added already in the IR to ensure the
  WCET analysis captures the correct function calls.
\end{romanenumerate}

The \prepphase starts with our \textsc{clang} plugin which collects the loop- and recursion-bound annotations,
described in Section~\ref{sec:llta-assumptions}.  It runs on the input
C source, before \llta is executed, and writes the collected bounds
to a JSON file that the \anaphase reads.
The JSON file is left out from \Cref{fig:llta-flow} for readability.

Next, the IR is emitted by \textsc{clang} and all available Runtime Libraries are linked at IR level using \textsc{llvm-link}, for afore mentioned reasons.
This IR file is enriched by running the following \llvm analysis and transformation passes using the \textsc{opt} tool, see \Cref{fig:llta-flow}:
\texttt{mem2reg}, \texttt{instcombine}, \texttt{loop-simplify}, \texttt{loop-rotate}, \texttt{indvars}.
Some of these passes have to be run before linking the IR, because they will erase unused code.
The library and ABI calls, e.g.: soft floats, will only be resolved and used in later compilation phases, so the IR optimizations tend to identify them as dead code.
For this reason we are also not able to have all passes executed by \llta.

The following passes in \llta to provide unified information for
CFG construction. Specifically six passes in this order: (i)
memory-intrinsic expansion, a pass implemented in \llta itself, to convert \texttt{memcpy}, \texttt{memmove} and
\texttt{memset} into bounded IR loops, (ii) scalar replacement of aggregates, making stack allocations accessible to scalar evolution, (iii) CFG simplification, for better loop recognition, (iv) loop rotation to
rewrite loops into the do/while formats preferred by \llvm's scalar-evolution
analysis, (v) loop-invariant code motion, moving code outside of loops, and (vi) induction-variable
simplification to simplify counters. Except for the first pass, which
\llta implements, all passes are \llvm's stock transformations
(\textsc{sroa}, \textsc{simplifycfg}, \textsc{loop-rotate},
\textsc{licm}, \textsc{indvars}), run unmodified.

The \prepphase further splits every BB at its call sites, so
that each call terminates its own block and the instructions after it
form the block to which the callee returns. This is going to be used
in \Cref{sec:llta-elf} to build the control flow graph.

The \prepphase ends by emitting the assembly of that MIR. The tool-chain of
the target platform assembles and
links it into the ELF binary, which is both the artifact flashed onto
the device and the second input of the \anaphase.

\subsection{The Analysis Phase}

This section explains how \llta takes the ELF binary and the \llvm IR, provided by the \prepphase, to derive a programs WCET during the \anaphase. It runs the same passes as the \prepphase right until assembly is emitted and thus holds the same information, only information from linking by the target tool-chain has to be reconstructed.
\subsubsection{The Analysis Graph}
\label{sec:llta-elf}

Since a worst-case path can cross function boundaries, the WCET
analysis needs one complete control flow graph for the whole program, denoted as the Analysis
Graph. The \anaphase builds it from the MIR. The graph contains one
node per machine BB, across all functions of the \llvm module. Call
and return edges connect these nodes across function boundaries. They
are the ordinary edges that the call-site split of
\Cref{sec:llta-prep} produced. A virtual entry node and a virtual exit
node anchor the graph at the function under analysis. Bounded
recursion is folded into bounded loop headers. Finally, every node
unreachable from the entry node is pruned.

However, this structure alone provides not enough information, as the compiler never sees the
final instruction addresses, the linker's relaxations, or the library
code that only the linker pulls in. The \anaphase lifts the
linked ELF binary into the graph, this is the \emph{linked-ELF
  facts} edge that runs from the ELF file into the \anaphase in
\Cref{fig:llta-flow}. That file is produced by the target tool chain. The \anaphase runs once the assembly emitted by the
\prepphase has been linked. It disassembles the binary and
aligns the decoded instruction stream with the MIR: each machine instruction is
re-encoded and matched against the binary. As a result, an address is
trusted only where the encodings agree. The linker may relax or
compress an instruction, which changes its size and thereby the
addresses of all following instructions. The alignment recognises such
instructions and keeps the two instruction streams aligned. A branch target read from the binary is accepted only when it agrees
with the resolved address of the target block. If part of a
function cannot be resolved, the bound is flagged unsound.

Four kinds of facts flow back into the analysis graph:
\begin{romanenumerate}
\item The real address of every instruction, on which the cache
  analysis and every placement-dependent cost depend.
\item The direction and target of every branch, from which per-edge
  costs are derived.
\item The addresses and sizes of data objects and sections.
\item The destination address of every direct call site is read from
  the linked bytes and compared against the entry address of the
  callee that the corresponding call edge in the graph points to. A
  disagreement means the linker redirected the call to a body other
  than the one analysed. In that case the bound is flagged unsound. A
  call site whose destination cannot be extracted from the bytes is
  counted as unverified rather than assumed to match.
\end{romanenumerate}

\begin{figure}[t]
	\centering
	\begin{minipage}[b]{0.40\linewidth}
		\begin{lstlisting}
void Multiply(matrix A,
    matrix B, matrix Res)
{
  register int Outer,
      Inner, Index;

  #pragma loop_bound(0, 20)
  for (Outer = 0;
       Outer < UPPERLIMIT;
       Outer++)
    #pragma loop_bound(0, 20)
    for (Inner = 0;
         Inner < UPPERLIMIT;
         Inner++)
    {
      Res[Outer][Inner] = 0;
      #pragma loop_bound(0, 20)
      for (Index = 0;
           Index < UPPERLIMIT;
           Index++)
        Res[Outer][Inner] +=
            A[Outer][Index]
            * B[Index][Inner];
    }
}
		\end{lstlisting}
		\vspace{3pt}
		\centering
		{\small (a) C source}
	\end{minipage}\hfill
	\begin{minipage}[b]{0.22\linewidth}
		\centering
		\includegraphics[width=\linewidth,height=100mm,keepaspectratio]{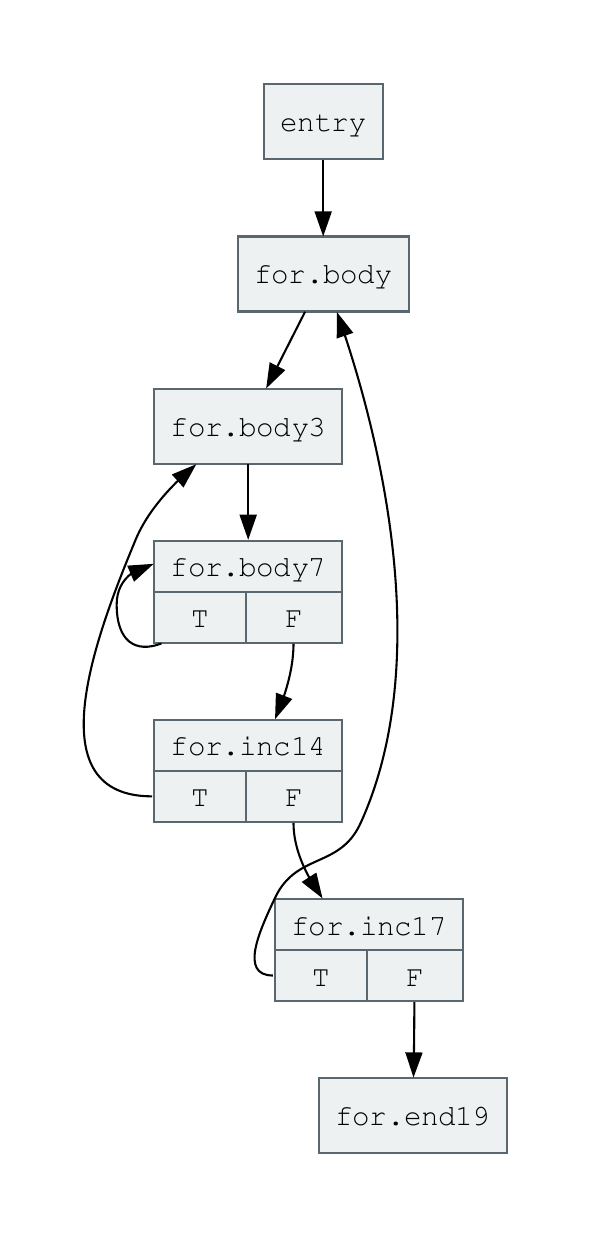}\\[3pt]
		{\small (b) LLVM~IR CFG}
	\end{minipage}\hfill
	\begin{minipage}[b]{0.36\linewidth}
		\centering
		\includegraphics[width=\linewidth,height=100mm,keepaspectratio]{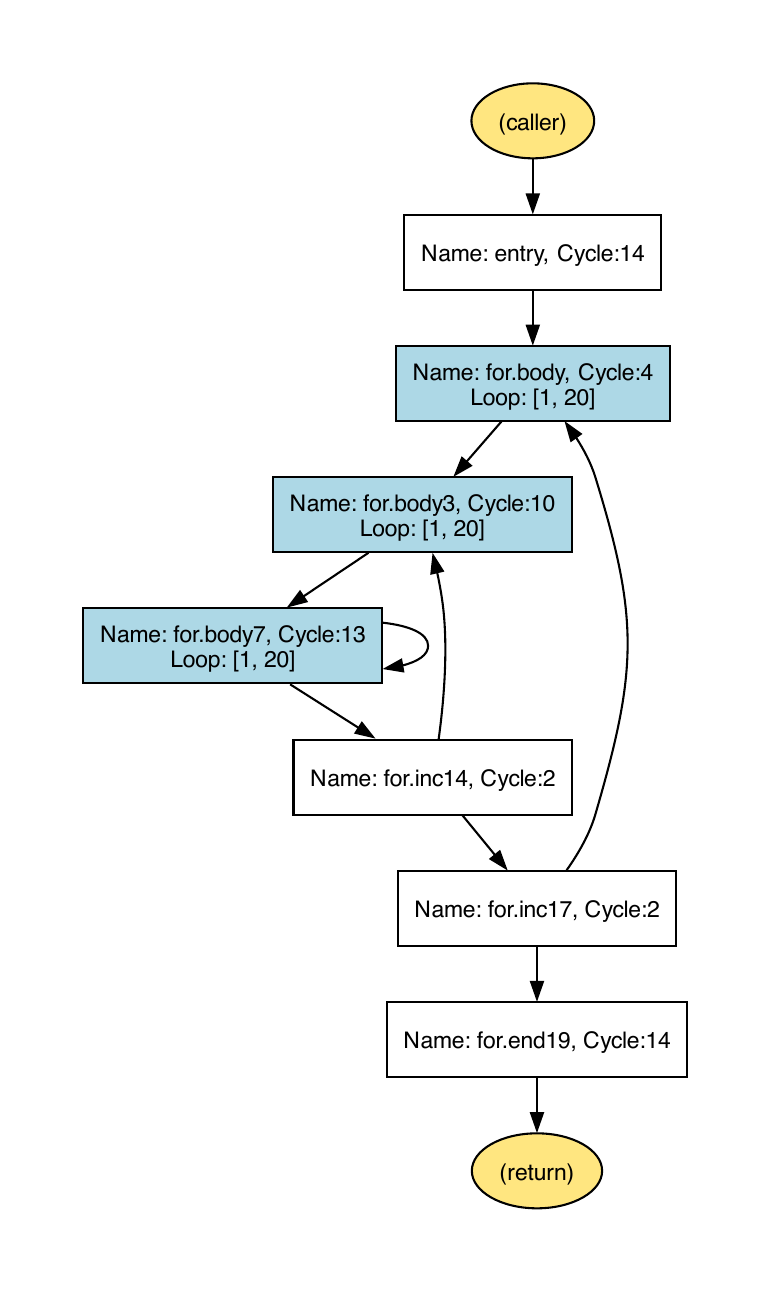}\\[3pt]
		{\small (c) LLTA machine CFG, ESP32-C6}
	\end{minipage}
	\caption{The \texttt{Multiply} function of \texttt{matmult}: (a) the C source with its loop-bound pragmas, (b) the LLVM~IR basic-block CFG, and (c) the machine-level CFG \llta analyses, here for the ESP32-C6, with per-block cycle costs and loop bounds.}
	\label{fig:cfg-compare}
\end{figure}

An example of using \llta to analyse the
\texttt{Multiply} function of \texttt{matmult} is shown in \Cref{fig:cfg-compare}. It shows the C source with
its loop-bound pragmas in~\Cref{fig:cfg-compare}(a), the
 CFG visible in the IR in~\Cref{fig:cfg-compare}(b),
and the machine-level graph \llta actually analyses in~\Cref{fig:cfg-compare}(c). Both~\Cref{fig:cfg-compare}(b)~and~(c)
are taken after the same IR-preparation passes. In~\Cref{fig:cfg-compare}(c), the collected information is integrated into the CFG. Here,
every block carries the cycle cost derived from its resolved machine
instructions, the three nested loop headers carry the bounds
\texttt{[1,20]} that become loop constraints of the path analysis, and
virtual entry and exit nodes anchor the flow. We note that the lower
bounds of the loops in~\Cref{fig:cfg-compare}(c) are $1$ instead of
the specified $0$ in the progmas in~\Cref{fig:cfg-compare}(a). This is
because the loop headers are contained in the BBs, which are executed at least once and SCEV corrects that.

\begin{figure}[t]
	\centering
	\includegraphics[height=110mm]{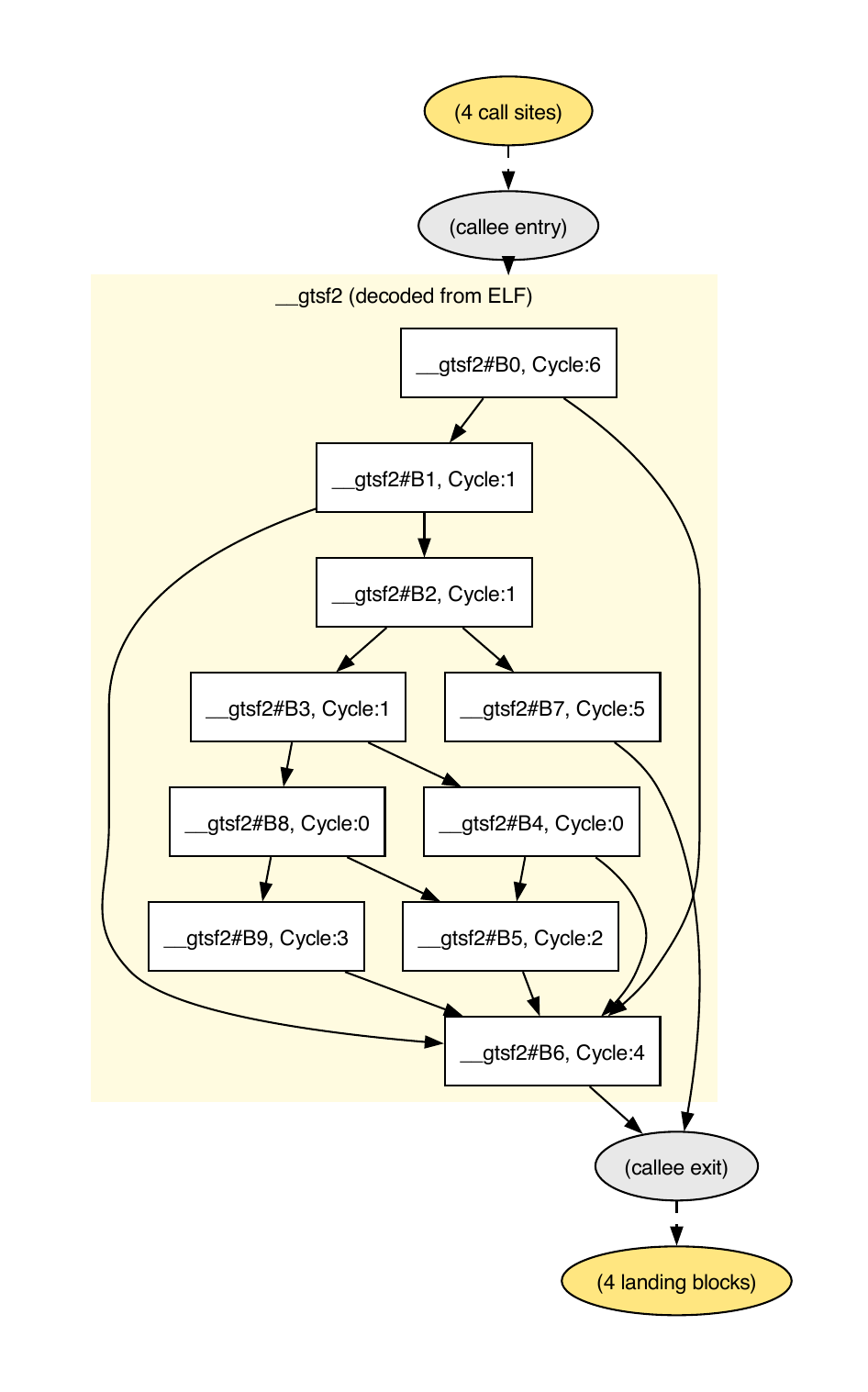}
	\caption{Library code recovered from the linked binary, in the analysis graph of the M\"alardalen \texttt{select} benchmark on the ESP32-C6. The \texttt{float} comparison \texttt{\_\_gtsf2} is one of the helpers that compiler-rt exposes as an alias. Hence, after the runtime library is linked in, the module holds no body for it and only the binary does. \llta decodes the helper's bytes, rebuilds the CFG shown here and stitches it into the analysis graph, where every decoded block carries the same measured per-instruction costs as compiled code and the cache and path analyses treat it like any other function. The callee is reached from four call sites and returns to the block following each of them, drawn here through virtual entry and exit nodes.}
	\label{fig:hybrid-call}
\end{figure}

\paragraph*{Library and ABI Calls}
\label{sec:abi-calls}
The analysed IR module does not define every function the program
calls. Runtime-library routines and ABI helpers, for example the
soft-float routines of compiler-rt, are only declared in the module
and receive their definition at link time. The worst-case path can run
through such callees, and, therefore, the analysis must cover them.
\llta does this by one of two routes:
\begin{romanenumerate}
\item \emph{The callee can be linked at IR level into the analysed module and thus \uline{holds} a body for the callee.}
  For this case, it is analysed as ordinary compiled code.  This
  brings the missing bodies into the module before the module reaches
  either phase, so that the \prepphase compiles them and the \anaphase
  analyses them. Both compiler-rt and llvm-libc are compiled to \llvm
  bitcode once, the \emph{Runtime Library (IR)} node of
  \Cref{fig:llta-flow}, and that bitcode is linked into the analysed
  module with \texttt{llvm-link -only-needed}, which imports exactly
  the transitively referenced bodies and nothing else.  The
  \texttt{llvm-link} happens at the IR level, before code generation.  Since
  the imported bodies run through the same preparation pipeline
  introduced in Section~\ref{sec:llta-prep}, the analysed module and
  the shipped module hold the same library code by construction.

\item \emph{The analysed IR module \uline{does not hold} a body for the
    callee and can only be linked by the target tool-chain.}  In this case, \llta decodes the callee's bytes and
  rebuilds its CFG.  This covers bodies that only the
  binary holds, e.g., ROM code, hand-written assembly, newlib's
  \texttt{atan}, and the helpers that compiler-rt exposes as an alias
  rather than as a definition, for which the runtime-library link
  imports no body.  \Cref{fig:hybrid-call} shows one of such callees: in
  the \texttt{select} benchmark, \texttt{\_\_lesf2} and
  \texttt{\_\_extendsfdf2} are imported and analysed as compiled code,
  while \texttt{\_\_gtsf2} and \texttt{\_\_ltsf2} are decoded. For
  this, \llta uses the decoder on the linked ELF file. From the decoded bytes it reconstructs the callee's CFG,
  as bottom-up analysers reconstruct control flow from
  binaries~\citep{Theiling2000}. The reconstructed blocks are then
  stitched into the analysis graph, with call and return edges
  connecting the call site to the decoded body. The result is one
  graph in which compiled code and library code are analysed alike.
  We further note that a callee whose
  reconstruction fails or whose reconstructed CFG contains a loop is
  refused and the corresponding bound is flagged unsound, since a loop in
  decoded machine code carries no bound.
\end{romanenumerate}

Once all this completes, the graph is frozen into an immutable Analysis
Graph, used to derive WCET as described in the following.

\subsubsection{Costs and Contexts}
\label{sec:llta-analyses}

The \anaphase charges costs to the nodes (basic blocks) of the Analysis Graph in two ways:
\begin{itemize}
\item When the execution of a node is not
  state-dependent, its execution cost is computed exactly once: every
  instruction is charged its latency from the target's cost table
  (\Cref{sec:targets}), a block's base cost is the sum, and call
  nodes carry the costs of the library and ABI callees of
  \Cref{sec:abi-calls}.
\item When the execution of a node is state-dependent, we utilize the
  instruction-cache must-analysis (\Cref{sec:background}) to analyze
  the execution cost. That is, a fetch is charged as a hit only where
  the must-analysis proves the line must be in the cache, and charge the full miss
  cost everywhere else. 
\end{itemize}

The contexts of the must-analysis are the VIVU contexts described in
\Cref{sec:background}, and the depth to which call strings are
truncated is configurable.

\subsubsection{Path Analysis}
\label{sec:path-analysis}

The \anaphase closes by deriving the worst-case execution path with
IPET (\Cref{sec:background}) over the whole-program graph. \llta uses the open-source HiGHS~\citep{Highs} solver, statically linked by \llta's build, in line with goal (G4). Therefore, integrating the solver is already done by \llta's build process. As the ILPs \llta produces are
small enough in our use case, we deem the performance of the solver sufficient.

\section{Target Devices}
\label{sec:targets}

In this section, we describe how we obtain the timing parameters of two
reference target devices, MSP430FR5994 and ESP32-C6.  The numbers reported in
this section are based on the calibrations which are consistent with our own
measurements and not a validated abstraction of the silicon.  We note again that
we make no claim about the correctness of either model.

\subsection{The ESP32-C6 Model}
\label{sec:target-esp32c6}

The ESP32-C6's application core is a 32-bit RISC-V (rv32imac) processor running
at 160\,MHz, single-issue, in-order and without a floating-point
extension. Code and data can live in three timing-relevant places:
the uncached internal high-performance (HP) SRAM, the external flash executed in place
through a cache, and the low-power (LP) SRAM. The model in
\Cref{tab:c6-latencies} was measured in the style of
nanoBench~\citep{Abel2020} and uops.info~\citep{Abel2019}. All rv32imac
instructions are extracted from \llvm's target description,
instantiated with safe operands, and validated by the \llvm
assembler. Each instruction is then timed on the device with the
on-chip cycle counter. Latency is measured over chains of 100
dependent instructions, and throughput over sequences of independent
ones. Where a cost varies with context or between builds,
the charged column takes the worst observation and rounds it up.

\begin{table}[t]
  \centering
  \caption{ESP32-C6 instruction and memory costs at 160\,MHz in
    cycles, as measured on the device and as charged by \llta.}
  \label{tab:c6-latencies}
  \footnotesize
  \begin{tabularx}{\linewidth}{>{\raggedright\arraybackslash}p{0.27\linewidth}
      rr>{\raggedright\arraybackslash}X}
    %\toprule
    class & measured & charged & note \\
    \midrule
    ALU, logic, shift, \textsc{mul} & 1.01 & 1 & \\ 
    \textsc{mulh} variants & 2.01 & 2 & \\ 
    \textsc{div}/\textsc{rem} & 10.01 & 10 &  \\ 
    sign/zero extension & 2.01 & 2 & \\ 
    atomics & 6.04 & 6 & Uncontended, on single core \\ 
    load SRAM & 3.00 & 3 &  High Performance SRAM\\
    load LP SRAM& 4.00 & $+1$ & Low Power (LP) SRAM \\
    cache miss & 346.5 & 347 & First-in, First-Out (FIFO)
      \\
    store & 1.02 & 1 & Measured with load contenders and without \\
    load & 38.0 & 38 & GPIO access \\
    store & 36.0 & 36 & GPIO access \\
    branch, predicted & 2.04 / 1.00 & 2 / 1 & Backward-taken /
      Forward-not-taken \\
    branch mispredicted & 3.51 & 4 & Forward-taken \\
    \textsc{jal}: direct jump & 3.00 & 3 & \\
    \textsc{jalr}: indirect jump, \texttt{ret} & 4.01 & 4 & \\
    %\bottomrule
  \end{tabularx}
\end{table}

The following three hardware properties of the core make these measured constants a sound
basis for static analysis.
\begin{romanenumerate}
\item First, the branch predictor applies static
logic, backward-taken, forward-not-taken (BTFN), with no
branch-target buffer and no return-address stack, so every
control-flow edge has a fixed cost independent of execution history
and the charged branch costs hold on every path. 

\item Second, the cache, 32\,KiB, 4-way with 32-byte cache lines and
  measured to replace in FIFO order, serves the single application core
  alone.

\item Third, \llta treats the core as if there is no pipeline execution. We observe, for example, a \textsc{div} instruction
  can complete one cycle earlier when it is followed by dependent
  instructions, in comparison to execution in isolation. The charged
  costs are the worst observations, so ignoring such overlap can only
  make a bound pessimistic.
\end{romanenumerate}
Cache hits and misses are decided
by the must-analysis of \Cref{sec:llta-analyses}, all other costs are
the static per-instruction and per-edge constants of
\Cref{tab:c6-latencies}.

\subsection{The MSP430 Model}
\label{sec:target-msp430}

The MSP430FR5994 is a 16-bit micro-controller built around FRAM, a
non-volatile memory that holds both code and data, next to a small
SRAM. The CPU executes instructions strictly sequentially, without
pipelining, and its family user's guide~\citep{SLAU445I} documents a
cycle count latency for every instruction, determined almost entirely by the
addressing modes. \Cref{tab:msp430-cycles} reproduces the documented
counts for the instructions. We validated the
documented latencies on silicon with assembly micro-benchmarks, and
every tested combination of instruction and addressing mode matches
the documented count exactly, with sequences costing exactly the sum
of their parts, so the model charges the documented counts.

\begin{table}[t]
  \centering
  \caption{Documented cycle counts of the MSP430X double-operand
    (Format~I) instructions by source and destination addressing
    mode, reproduced from the family user's
    guide~\citep{SLAU445I}. Source and Destination are register types of the operands.}
  \label{tab:msp430-cycles}
  \footnotesize
  \begin{tabular}{lrrrrr}
    & \multicolumn{5}{c}{destination} \\
    source & \texttt{Rm} & \texttt{PC} & \texttt{x(Rm)} & \texttt{EDE} & \texttt{\&EDE} \\
    \midrule
    \texttt{Rn}            & 1 & 3 & 4 & 4 & 4 \\
    \texttt{@Rn}           & 2 & 4 & 5 & 5 & 5 \\
    \texttt{@Rn+}          & 2 & 4 & 5 & 5 & 5 \\
    \texttt{\#N}           & 2 & 3 & 5 & 5 & 5 \\
    \texttt{x(Rn)}         & 3 & 5 & 6 & 6 & 6 \\
    \texttt{EDE}           & 3 & 5 & 6 & 6 & 6 \\
    \texttt{\&EDE}         & 3 & 5 & 6 & 6 & 6 \\
  \end{tabular}
\end{table}

The FRAM of MSP430FR5994 keeps up with the CPU only up to
8\,MHz. Above that, the device inserts wait states and hides part of
their cost behind a small instruction-fetch cache. The timing properties of this
cache are not documented. We therefore reverse-engineered it with
timing probes. The cache holds 2 sets of 2 ways with 8-byte lines and
uses a least-recently-used (LRU) replacement policy. A miss stalls the
fetch for a 15-cycle line fill. \llta analyses this cache with the
must-analysis of \Cref{sec:llta-analyses}. By default, the analysis
assumes an unknown replacement policy, which is the conservative
choice. Since our measurements indicate LRU replacement, an LRU
must-analysis can be selected instead.

\paragraph*{Limitations of MSP430} 
The remaining limitations of the target stem from \llvm. Its MSP430
backend implements only the older 16-bit instruction set
architecture (ISA), while the linked libgcc and newlib helpers are
20-bit MSP430X code. This constrains \llta to decode helper bodies from the
binary, which is possible on the ESP32-C6 (\Cref{sec:abi-calls}).
Therefore, such calls have to be pre-measured and added to the analysis graph as static costs like an instruction.
Due to this black box method on the MSP430, the memory analysis becomes infeasible as memory accesses at such call sites are entirely unknown.

\section{Case Study}
\label{sec:llta-eval}
% Maelardalen measured-vs-bound table (tab:mael-wcet). Inlined from
% mael-wcet-table.tex because Springer forbids \input. Rows regenerate
% from the LLTA repo:
%   python3 tests/hardware_regression_test.py msp430  --format latex
%   python3 tests/hardware_regression_test.py riscv32 --format latex
\begin{table}[t]
  \centering
  \footnotesize
  \setlength{\tabcolsep}{5pt}
  \caption{Measured maximal observed cycles versus the WCET bound computed by \llta for the M\"alardalen benchmarks on the MSP430FR5994 (8\,MHz) and the ESP32-C6 (160\,MHz). A dagger ($\dagger$) marks a bound \llta reports unsound, and \textbf{bold} marks a bound below the measured cycle count.}
  \label{tab:mael-wcet}
  \begin{tabular}{lrrrr}
    %\toprule
    & \multicolumn{2}{c}{MSP430FR5994 (8\,MHz)} & \multicolumn{2}{c}{ESP32-C6 (160\,MHz)}\\
    \cmidrule(lr){2-3}\cmidrule(lr){4-5}
    Benchmark & Measured & LLTA & Measured & LLTA\\
    \midrule
    adpcm          & -- & -- & 187833 & 308321\\
    bs             & 134 & 186 & 99 & 162\\
    bsort100       & 156119 & 326359 & 83272 & 256785\\
    cnt            & 18243 & 83547 & 4608 & 6278\\
    compress       & 25138 & 1072607 & 4811 & 195207\\
    cover          & 3124 & 3483 & 2196 & 3166\\
    crc            & 39775 & 100588 & 28280 & 61521\\
    des            & -- & 249082 & 1075274 & 1315679\\
    duff           & 1188 & 1575 & 846 & 951\\
    edn            & 521583 & 2759325 & 51018 & 74289\\
    expint         & 30996 & 623525 & 3377 & 4799\\
    fac            & 858 & 12108 & 402 & 1075\\
    fdct           & 5156 & 33759 & 1867 & 2266\\
    fft1           & 2355850 & \uwcet{1664666$\dagger$} & 49855 & 2657149\\
    fibcall        & 269 & 327 & 187 & 309\\
    fir            & -- & -- & 210758 & 311566\\
    insertsort     & 1036 & 1928 & 562 & 1666\\
    janne\_complex & 739 & 30777 & 261 & 2257\\
    jfdctint       & 11263 & 84380 & 2288 & 2590\\
    lcdnum         & 248 & 446 & 197 & 557\\
    lms            & 126381467 & 156597611$\dagger$ & 7068635 & 56686608\\
    ludcmp         & -- & -- & 40634 & 195505\\
    matmult        & 471406 & 3408571 & 99815 & 143211\\
    minver         & 479406 & \uwcet{27211$\dagger$} & 14524 & 57382\\
    ndes           & 188134 & 319567 & 57308 & 79766\\
    ns             & 7695 & 9543 & 4874 & 6285\\
    nsichneu       & 16939 & 33448 & 5160 & 15222\\
    prime          & 4871 & 563713 & 12220 & 25944\\
    qsort-exam     & 67407 & 1022962$\dagger$ & 3232 & 1185475\\
    qurt           & 807038 & \uwcet{14961$\dagger$} & 19857 & 132068\\
    recursion      & 4701 & 6916 & 2844 & 7101\\
    select         & 944943400 & \uwcet{527119$\dagger$} & 2433 & 596399\\
    sqrt           & 103247 & 123544$\dagger$ & 6752 & 37271\\
    st             & -- & -- & 3859113 & 6878474\\
    statemate      & 1543 & 3230 & 1546 & 3195\\
    ud             & -- & -- & 3749 & 25577\\
    whet           & 1582980252 & \uwcet{11303643$\dagger$} & 76439094 & 288158196$\dagger$\\

    %\bottomrule
  \end{tabular}
\end{table}

We evaluate \llta on the M\"alardalen suite~\citep{Gustafsson2010} using an MSP430FR5994 Launchpad and an ESP32-C6 DevKit.
We measure the execution cycles of
each flashed binary on the device itself, and compare them against
the bounds \llta computes for the same binary.

\Cref{tab:mael-wcet}
shows the output of \llta and the measurements for each benchmark on both devices. Each benchmark runs
with its single input, which is part of the benchmark itself, exactly
five times, and the table reports the longest run. A dagger
($\dagger$) marks a bound \llta flags as unsound, and bold marks a
bound below the measured count.

On both devices, every bound that \llta reports as sound
upper-bounds the measured execution. Since every run executes the
same single input, the measurements do not guarantee worst-case
coverage. However, as the bounds for the actual binary that is flashed onto the devices, \llta fulfills goal G2.

In line with goal G3, \llta reports several benchmark results as unsound bounds.
On the MSP430FR5994, eight benchmarks use library routines which \llta flags~($\dagger$), as described in \Cref{sec:abi-calls}. Five of
these flagged bounds are indeed exceeded by the measurement.
On the ESP32-C6 the same routines are analysed as \llvm IR
(\Cref{sec:abi-calls}) and, therefore, \llta can bound them.
Only \texttt{whet} remains flagged, since it calls routines that exist
only in the linked binary and that \llta's decoding route refuses
(\Cref{sec:abi-calls}).

Not every benchmark is usable on the MSP430FR5994, and the table
leaves their cells empty. The device holds 8\,KiB of SRAM, but the
16-bit instruction format \llta supports can address only 4\,KiB of
it. Four benchmarks exceed that limit at link time. \texttt{adpcm}
could not be compiled due to a code-generation defect in LLVM~20's
MSP430 backend. \texttt{des} analyses to a sound bound, listed in the
table, but hangs on the Launchpad, leaving no measurement to compare
against.

\section{LLTA for Teaching}
\label{sec:exercises}
In this section, we sketch how \llta can be used for teaching WCET analyses in three scenarios that can be used in an RTOS or general real-time systems course:

\begin{itemize}
  \item \textbf{Determining WCET bounds}: \llta can be used to provide bounds for  real-time tasks that students program. Either a sound WCET bound is returned for the analysed task, or an unsound bound is returned in case the students write the task in a way where a safe bound cannot be given. Students can also be instructed to revise the C code until \llta returns a safe bound, which can be an assignment for teaching students analysable coding styles. The WCET bound given by \llta can also be utilized to study how the WCET of a task can be improved by code optimization or optimizations via the compiler.
\item \textbf{Interplay of WCET and Computer Architecture}: To
  estimate the execution time of an instruction, it is unavoidable to
  investigate the underlying computer architecture. Students can
  derive a cost table for a core with an \llvm backend, and measure
  the timing of instructions, and compare their findings against the
  existing specifications (if they exist). This can start from the
  models derived in \Cref{sec:targets}. It can
  also be extended to study the impact of the cache-replacement-policy
  model on the WCET. The cache analysis of \Cref{sec:llta-analyses} is based on a modifiable cache configuration and replacement policy. Students can implement a different policy and observe its effect on the bounds of \Cref{sec:llta-eval}.
\item \textbf{Advanced WCET Analyses}: While \llta provides simple WCET analyses, advanced
  projects can also incorporate more complex WCET analyses. For example, a loop-bound analysis
  can be further included to safely analyse bounds on trip counts that
  scalar evolution cannot derive.
\end{itemize}

\section{Conclusion and Outlook}
\label{sec:conclusion}

This paper presents \llta, a tool chain for WCET analyses on COTS hardware, designed for simplicity and usage especially
in teaching environments or newcomers of worst-case timing analysis.
For this, \llta
builds solely upon open-source tools such that there is no need to maintain licensing or rely on any university program of commercial tools.
Additionally, \llta injects its analysis passes at the hook points of an unmodified
\llvm, i.e., there is no patched compiler or tool chain to maintain.
We provide a tool chain that is simple (G1), runs on real COTS hardware (G2), reports results that are unsound due to programming or device constraints (G3), and is available open-source without licensing complications (G4).

Our research group has developed some tools, that are highly relevant
to the engineering process of real-time systems. For example,
TEACH~\citep{Teach} supplies a deployable setup for hardware running FreeRTOS alongside remote
access to it, and a body of assignments, although it was originally
designed for ESP32-S3 back to 2023, the latest version has been
migrated to ESP32-C6 to cover RISC-V based devices. We will soon integrate \llta and provide
materials for RTOS hands-on engineering. In addition,
SLIME~\citep{hakert2026slime} analyses the binary code of an RTOS to
extract the timing information of the real-time tasks. SLIME provides the period $T_i$ of a real-time task $\tau_i$ alongside potentially the suspension
interval $S_i$, while \llta supplies the WCET $C_i$ of every task $\tau_i$. By combining the features of \llta and SLIME, we can ensure that the analyses are done directly on the binary
program flashed onto the device.

\backmatter

\bmhead{Supplementary information}

LLTA, the two target models and the build integration described in this article are available as open source at \url{https://github.com/tu-dortmund-ls12-rt/LLTA}.

\section*{Declarations}

\bmhead{Funding}
This work has been supported by European Research Council (ERC) Consolidator Award 2019, as part of PropRT (Number 865170), and by Deutsche Forschungsgemeinschaft (DFG), as part of One-Memory (405422836), and by DFG Priority Program “Disruptive Memory Technologies” (SPP 2377) as part of the project “ARTS-NVM” (502308721).

\bmhead{Conflict of interest}
The authors declare no competing interests.

\bmhead{Ethics approval and consent to participate}
Not applicable.

\bmhead{Consent for publication}
Not applicable.

\bmhead{Data availability}
Not applicable.

\bmhead{Materials availability}
Not applicable.

\bmhead{Code availability}
The source code of \llta{} is available at \url{https://github.com/tu-dortmund-ls12-rt/LLTA}.

%\bmhead{Author contribution}

%\bmhead{Acknowledgements}

\bibliography{real-time}

\end{document}